\documentclass[12pt]{article}
\usepackage{epsfig}

\def\beq{\begin{equation}}
\def\eeq#1{\label{#1}\end{equation}}
\def\eeqn{\end{equation}}

\def\beqa{\begin{eqnarray}}
\def\eeqa#1{\label{#1}\end{eqnarray}}
\def\eeqan{\end{eqnarray}}

\let\bar=\overbar

\def\Dslash{\not{\hbox{\kern-4pt $D$}}}
\def\dslash{\not{\hbox{\kern-2pt $\del$}}}

\def\msb{{\bar{\ssstyle M \kern -1pt S}}}

\def\Title#1{\begin{center} {\Large {\bf #1} } \end{center}}

\begin{document}

\noindent
Proceedings of CKM 2012,\\
the 7$^{th}$ International Workshop on the CKM Unitarity Triangle,\\
University of Cincinnati, USA, 28 September - 2 October 2012

\bigskip\bigskip

\Title{Determination of $|V_{cb}|$ from Inclusive Decays $B\to
  X_c\ell\nu$ using a Global Fit}

\bigskip\bigskip


\begin{raggedright}
{\it Christoph Schwanda\index{Schwanda, C.}\\
Institute of High Energy Physics\\
Austrian Academy of Sciences\\
A-1050 Vienna, AUSTRIA}
\bigskip\bigskip
\end{raggedright}

\section{Introduction}

In this article, we review the theory and the experimental status of
inclusive semileptonic $B$~meson decays $B\to X_c\ell\nu$. Based on
these inputs, we present the latest determination of the magnitude of
the Cabibbo-Kobayashi-Maskawa matrix element
$|V_{cb}|$~\cite{Kobayashi:1973fv} and of the $b$-quark mass $m_b$,
obtained by the Heavy Flavor Averaging Group (HFAG).

\section{Theory}

The theoretical tool for calculating the inclusive semileptonic $B$~decay
width $\Gamma(B\to X_c\ell\nu)$ is the Operator Product Expansion
(OPE)~\cite{Benson:2003kp,Bauer:2004ve},
\begin{equation}
  \Gamma(B\to
  X_c\ell\nu)=\frac{G^2_Fm^5_b}{192\pi^3}|V_{cb}|^2\left(1+\frac{c_5(\mu)\,\langle
    O_5\rangle(\mu)}{m^2_b}+\frac{c_6(\mu)\,\langle
    O_6\rangle(\mu)}{m^3_b}+\mathcal{O}(\frac{1}{m^4_b})\right)~. \label{eq:1}
\end{equation}
At the leading order in $1/m_b$, this expression coincides with the
parton model result, {\it i.e.}, the decay width of a free
$b$-quark. At higher orders appear Wilson
coefficients ($c_5, c_6$) multiplying expectation values of local
operators ($\langle O_5\rangle, \langle O_6\rangle$). The coefficients
$c_5$ and $c_6$ contain the perturbative QCD physics while the local
operators are non-perturbative objects.

Equation \ref{eq:1} assumes parton-hadron duality, which implies that the
hadronic parameters $\langle O_5\rangle$ and $\langle O_6\rangle$ do
not depend on the final state of the decay. Thus, they also appear in
OPEs for other inclusive $B$~meson observables and can be measured in
experiments. This entire procedure is referred to as a global fit.

The inclusive observables used in this analysis are the
(truncated) moments of the lepton energy $E_\ell$ (in the $B$~rest
frame) and the $m^2_X$ spectra in $B\to
X_c\ell\nu$, where $m^2_X$ is the invariant mass squared of the hadronic
system $X_c$ accompanying the lepton-neutrino pair. The lepton energy
moments are defined as
\begin{equation}
  \langle
  E^n_\ell\rangle_{E_\mathrm{cut}}=\frac{R_n(E_\mathrm{cut})}{R_0(E_\mathrm{cut})}~,
  \quad
  R_n(E_\mathrm{cut})=\int_{E_\ell>E_\mathrm{cut}}E^n_\ell\frac{d\Gamma}{dE_\ell}dE_\ell~,
\end{equation}
where $E_\mathrm{cut}$ is the lower lepton energy threshold and
$d\Gamma/dE_\ell$ is the partial semileptonic width as a function of
the lepton energy. The hadronic mass moments are
\begin{equation}
  \langle
  m^{2n}_X\rangle_{E_\mathrm{cut}}=\frac{S_n(E_\mathrm{cut})}{S_0(E_\mathrm{cut})}~,
  \quad
  S_n(E_\mathrm{cut})=\int_{E_\ell>E_\mathrm{cut}}m^{2n}_X\frac{d\Gamma}{dm^2_X}dm^2_X~.
\end{equation}
Also here, the integration is over the $B\to X_c\ell\nu$ phase space
restricted by the requirement $E_\ell>E_\mathrm{cut}$.

OPE calculations of the semileptonic width and these moments have been
obtained in two theoretical frameworks, referred to by the
name of the renormalization scheme used for the quark masses. The
calculations in the \emph{kinetic scheme} are now available at
next-to-next-to-leading (NNLO) order in
$\alpha_s$~\cite{Benson:2003kp,Gambino:2011cq}. At leading order in
the OPE, the non-perturbative parameters are the quark masses $m_b$
and $m_c$. At $\mathcal{O}(1/m^2_b)$ ($\mathcal{O}(1/m^3_b)$) the
parameters are $\mu^2_\pi$ and $\mu^2_G$ ($\rho^3_D$, $\rho^3_{LS}$).
Independent expressions are available in the
\emph{1S scheme}~\cite{Bauer:2004ve}. Here, the long-distance
parameters are $m_b$ at leading order, $\lambda_1$ and $\lambda_2$
at $\mathcal{O}(1/m^2_b)$ and $\rho_1$, $\tau_{1-3}$ at
$\mathcal{O}(1/m^3_b)$. Note that the numerical values of the quark
masses in the two schemes cannot be compared directly due to their
different definitions. The expressions used here include power
corrections up to $\mathcal{O}(1/m^3_b)$ though higher orders have
already been calculated~\cite{Mannel:2010wj}.

\section{Experiment}

The most precise measurements of partial semileptonic branching
fractions and moments in $B\to X_c\ell\nu$ are obtained by the
Belle~\cite{Urquijo:2006wd,Schwanda:2006nf} and BaBar
collaborations~\cite{Aubert:2009qda} analyzing 152 and 232 million
$\Upsilon(4S)\to B\bar B$~events, respectively. These studies proceed as
follows: first, the decay of one $B$~meson in the event is fully
reconstructed in a hadronic mode ($B_\mathrm{tag}$). Then, the
semileptonic decay of the second $B$~meson ($B_\mathrm{sig}$) is
identified by searching for a charged lepton among the remaining
particles in the event.

The observed $E_\ell$ and $m^2_X$ spectra are distorted by resolution
and acceptance effects. Belle corrects for this by unfolding the
observed spectra using the Singular Value Decomposition (SVD)
algorithm~\cite{Hocker:1995kb} and measures the energy moments
$\langle E^k_\ell\rangle$ for $k=0,1,2,3,4$ and minimum lepton
energies ranging from 0.4 to 2.0~GeV. Moments of the hadronic
mass~$\langle m^k_X\rangle$ are measured for $k=2,4$ and minimum
lepton energies from 0.7 to 1.9~GeV. BaBar applies
a set of linear corrections, which depend on 
the charged particle multiplicity of the $X$~system, the normalized
missing mass, $E_\mathrm{miss}-p_\mathrm{miss}$, and the lepton
momentum. In this way, BaBar measures the moments of the hadronic mass
spectrum up to $\langle m^6_X\rangle$ for minimum lepton energies
ranging from 0.8 to 1.9 GeV. In Ref.~\cite{Aubert:2009qda} the
earlier measurement of the lepton energy moments in $B\to
X_c\ell\nu$~\cite{Aubert:2004td} is updated using new branching
fraction measurements for background decays and an improved evaluation
of systematic uncertainties.

All measurements used by HFAG for
determining $|V_{cb}|$ inclusive are listed in Table~\ref{tab:1}. The
only external input for the fit is the average lifetime~$\tau_B$ of
neutral and charged $B$~mesons, taken to be $(1.582\pm
0.007)$~ps~\cite{Beringer:1900zz}.
\begin{table}
\caption{Experimental inputs for the global analysis of $B\to
  X_c\ell\nu$. $n$ is the order of the moment, $c$ is the
  threshold value in GeV. In total, there are 29 measurements from
  BaBar, 25 measurements from Belle and 12 from other
  experiments.} \label{tab:1}
\begin{center}
\resizebox{0.99\textwidth}{!}{
\begin{tabular}{l|l|l|l}
  \hline \hline
  Experiment
  & Hadron moments $\langle M^n_X\rangle$
  & Lepton moments $\langle E^n_\ell\rangle$
  & Photons moment $\langle E^n_\gamma\rangle$\\
  \hline
  BaBar & $n=2$, $c=0.9,1.1,1.3,1.5$ & $n=0$, $c=0.6,1.2,1.5$ & $n=1$,
  $c=1.9,2.0$\\
  & $n=4$, $c=0.8,1.0,1.2,1.4$ & $n=1$, $c=0.6,0.8,1.0,1.2,1.5$ & $n=2$,
  $c=1.9$~\cite{Aubert:2005cua,Aubert:2006gg}\\
  & $n=6$, $c=0.9,1.3$~\cite{Aubert:2009qda} & $n=2$, $c=0.6,1.0,1.5$
  & \\
  & & $n=3$, $c=0.8,1.2$~\cite{Aubert:2009qda,Aubert:2004td} & \\
  \hline
  Belle & $n=2$, $c=0.7,1.1,1.3,1.5$ & $n=0$, $c=0.6,1.0,1.4$ & $n=1$,
  $c=1.8,1.9$\\
  & $n=4$, $c=0.7,0.9,1.3$~\cite{Schwanda:2006nf} & $n=1$,
  $c=0.6,0.8,1.0,1.2,1.4$ & $n=2$, $c=1.8,2.0$~\cite{Limosani:2009qg}\\
  & & $n=2$, $c=0.6,1.0,1.4$ & \\
  & & $n=3$, $c=0.8,1.0, 1.2$~\cite{Urquijo:2006wd} & \\
  \hline
  CDF & $n=2$, $c=0.7$ & & \\
  & $n=4$, $c=0.7$~\cite{Acosta:2005qh} & & \\
  \hline
  CLEO & $n=2$, $c=1.0,1.5$ & & $n=1$, $c=2.0$~\cite{Chen:2001fja}\\
  & $n=4$, $c=1.0,1.5$~\cite{Csorna:2004kp} & & \\
  \hline
  DELPHI & $n=2$, $c=0.0$ & $n=1$, $c=0.0$ & \\
  & $n=4$, $c=0.0$~\cite{Abdallah:2005cx} & $n=2$, $c=0.0$ & \\
  & & $n=3$, $c=0.0$~\cite{Abdallah:2005cx} & \\
  \hline \hline
\end{tabular}
}
\end{center}
\end{table}

\section{Results for $|V_{cb}|$ inclusive and $m_b$}

By fitting the measurements in Table~\ref{tab:1} to the OPE
expressions of the semileptonic width and of the $B\to
X_c\ell\nu$~moments, properly
accounting for correlations in the theory expressions and experimental
data, $|V_{cb}|$, the $b$-quark mass and the other hadronic parameters
are obtained. The moments in $B\to X_c\ell\nu$ are sufficient for
determining $|V_{cb}|$ but measure the $b$-quark mass only to about
50~MeV precision.
Therefore, additional constraints are introduced: the photon
energy moments in $B\to X_s\gamma$, or a precise constraint on the
$c$-quark mass.
For the former, calculations of the $B\to X_s\gamma$~moments are
available both in the kinetic~\cite{Benson:2004sg} and the
$1S$~scheme~\cite{Bauer:2004ve}. For the latter, HFAG uses the $c$-quark
mass calculated in Ref.~\cite{Dehnadi:2011gc}, $m_c^{\overline{\rm
    MS}}(3~{\rm GeV})=(0.998\pm 0.029)$~GeV, obtained using low-energy
sum rules. Note that the $c$-quark
mass constraint cannot be applied
in the $1S$~scheme as the $1S$ expressions do not depend on this
parameter.

The results of the HFAG analysis in the kinetic scheme are given in
Table~\ref{tab:2} for both choices of the additional constraint. Note
the excellent agreement in the $b$-quark mass, which is known from
this study to almost 20~MeV precision. The relative
uncertainty in $|V_{cb}|$ is about 1.7\%. Table~\ref{tab:3} contains
the results in the $1S$~scheme for $B\to X_c\ell\nu$ only and using
the $B\to X_s\gamma$~constraint. The central value of $|V_{cb}|$ is in
excellent agreement with the kinetic scheme analysis. Due to a more
aggressive error estimate, the relative precision here is 1.1\%. The
full result for all hadronic parameters and the entire
correlation matrix is given in Ref.~\cite{Amhis:2012bh}.
\begin{table}
\caption{Global fit results in the kinetic scheme for different
  constraints.} \label{tab:2}
\begin{center}
\begin{tabular}{c|c|c|c|c}
  \hline \hline
  Constraint & $|V_{cb}|$ (10$^{-3}$) & $m_b^{\rm kin}$ (GeV) &
  $\mu^2_\pi$ (GeV$^2$) & $\chi^2/$d.o.f.\\
  \hline
  $B\to X_s\gamma$ & $41.94\pm 0.43_{\rm fit}\pm 0.59_{\rm th}$ &
  $4.574\pm 0.032$ & $0.459\pm 0.037$ & $27.0/(66-7)$\\
  $m_c^{\overline{\rm MS}}(3~{\rm GeV})$ & $41.88\pm 0.44_{\rm fit}\pm
  0.59_{\rm th}$ & $4.560\pm 0.023$ & $0.453\pm 0.036$ &
  $33.4/(55-7)$\\
  \hline \hline
\end{tabular}
\end{center}
\end{table}
\begin{table}
\caption{Global fit results in the $1S$ scheme for different
  constraints.} \label{tab:3}
\begin{center}
\begin{tabular}{c|c|c|c|c}
  \hline \hline
  Constraint & $|V_{cb}|$ (10$^{-3}$) & $m_b^{1S}$ (GeV) &
  $\lambda_1$ (GeV$^2$) & $\chi^2/$d.o.f.\\
  \hline
  $B\to X_s\gamma$  & $41.96\pm 0.45$ & $4.691\pm 0.037$ &
  $-0.362\pm 0.067$ & $23.0/(66-7)$\\
  None & $42.37\pm 0.65$ & $4.622\pm 0.085$ & $-0.412\pm
  0.084$ & $13.7/(55-7)$\\
  \hline
\end{tabular}
\end{center}
\end{table}

\end{document}